\documentclass[aps,prl,reprint,floatfix,superscriptaddress]{revtex4-1}
\usepackage[english]{babel}
\usepackage[T1]{fontenc}
\usepackage{bbold}
\usepackage{epsfig}
\usepackage{amsmath}
\usepackage{hyphenat}
\usepackage{placeins}
\usepackage{color}

\usepackage{array}
\newcolumntype{C}[1]{>{\centering\arraybackslash}m{#1}}
\usepackage{overpic}

\DeclareMathOperator{\Hamil}{\mathcal{H}}  

\begin{document}

\title{Emergence of a 2d macro-spin liquid in a highly frustrated 3d quantum magnet}
\author{Tycho S. Sikkenk}
\affiliation{Institute for Theoretical Physics and Center for Extreme Matter and Emergent Phenomena, Utrecht University, Leuvenlaan 4, 3584 CE Utrecht, The Netherlands}
\author{Kris Coester}
\affiliation{Lehrstuhl f\"ur Theoretische Physik I, Otto-Hahn-Str. 4, TU Dortmund, D-44221 Dortmund, Germany}
\author{Stefan Buhrandt}
\affiliation{Institute for Theoretical Physics and Center for Extreme Matter and Emergent Phenomena, Utrecht University, Leuvenlaan 4, 3584 CE Utrecht, The Netherlands}
\author{Lars Fritz}
\affiliation{Institute for Theoretical Physics and Center for Extreme Matter and Emergent Phenomena, Utrecht University, Leuvenlaan 4, 3584 CE Utrecht, The Netherlands}
\author{Kai P. Schmidt}
\affiliation{Lehrstuhl f\"ur Theoretische Physik I, Otto-Hahn-Str. 4, TU Dortmund, D-44221 Dortmund, Germany}

\begin{abstract}
The classical Ising model on the frustrated 3d swedenborgite lattice has disordered spin liquid ground states for all ratios of inter- and intra-planar couplings. Quantum fluctuations due to a transverse field give rise to several exotic quantum phenomena. In the limit of weakly coupled Kagom\'e layers we find a 3d version of disorder by disorder. For large out-of-plane couplings 1d macro-spins are formed which realize a disordered macro-spin liquid on an emerging triangular lattice. Signatures of this dimensional reduction are also found in critical exponents of the quantum phase transition out of the fully polarized phase into the macro-spin liquid displaying quantum criticality typical for 2d quantum systems.       
\end{abstract}

\maketitle

{\it{Introduction:}} Geometrical frustration in magnetic systems can give rise to a multitude of exotic classical and quantum phases. An analysis of the classical limit often reveals a large degeneracy in the ground-state manifold, which can be lifted by quantum or thermal fluctuations thereby selecting an ordered state. This phenomenon is conventionally referred to as 'order by disorder'. A more exotic version of degeneracy lifting is called 'disorder by disorder' \cite{Moessner2001,Moessner2000,Priour2001,Powalski2013,Coester2013}: out of multiple classical ground states a disordered state is selected. Paradigmatic examples for either of the two phenomena can be found in two dimensions for the transverse field Ising model (TFIM), on the triangular and the Kagom\'e lattices, respectively. In both cases a polarized phase is found at high transverse fields. On the triangular lattice an infinitesimal transverse field is sufficient to select an ordered state out of the degenerate classical manifold, thereby providing an example of order by disorder. The selected state is the so-called $\sqrt{3}\times\sqrt{3}$-state which maximizes the number of flippable spins. The two phases are connected via a second order phase transition in the 3d XY universality class \cite{Moessner2001,Isakov2003,Powalski2013}. In contrast, in the case of the Kagom\'e lattice any finite transverse field selects the disordered polarized phase which does not break any symmetry, providing an example of disorder by disorder \cite{Moessner2001,Powalski2013}. 
% Figure 1: Phase diagram of the TFIM on the Swedenborgite lattice
%%%%%%%%%%%%%%%%%%%%%%%%%%%%%%%%%%%%%%%%%%%%%%%%%%%%%%%%%%%%%%%%%%%%%%%%%%%%%%%%%%%%%%%%%%%%%%%%%%%%%%%%%%%%%%%%%%%%%%%%%%%%%%%%
\begin{figure} [h!]
 \includegraphics[width=0.45\textwidth]{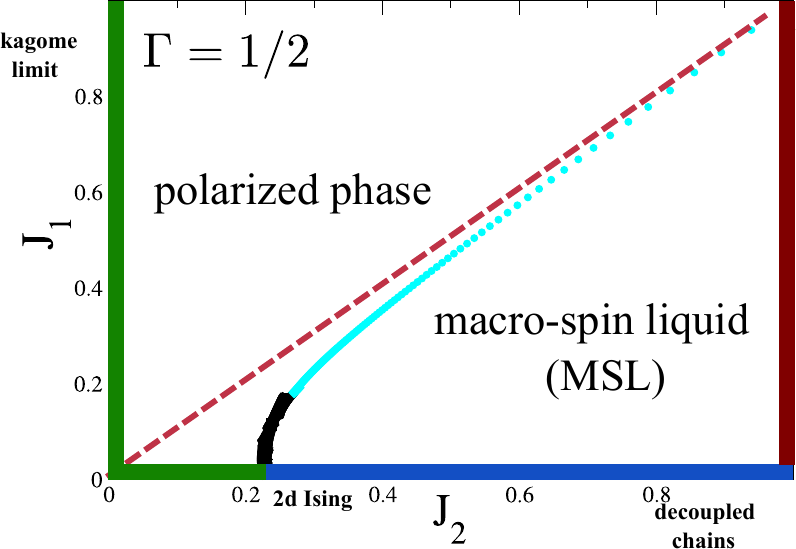}
 \caption{Phase diagram of the transverse field Ising model on the swedenborgite lattice as a function of $J_1$ and $J_2$ for the field $\Gamma=1/2$. Black and cyan symbols correspond to different extrapolation schemes for the breakdown of the polarized phase obtained from different extrapolations of the high-field gap obtained by perturbative continuous unitary transformations (pCUTs) (for more details see Supplemental Material Fig.~\ref{fig:dlogpade}). The dashed line denotes $J_1=J_2$, the classical phase transition line between two different types of spin liquids. The blue line denotes an ordered phase in the strictly one-dimensional limit.}
 \label{fig:pd}
\end{figure}
%%%%%%%%%%%%%%%%%%%%%%%%%%%%%%%%%%%%%%%%%%%%%%%%%%%%%%%%%%%%%%%%%%%%%%%%%%%%%%%%%%%%%%%%%%%%%%%%%%%%%%%%%%%%%%%%%%%%%%%%%%%%%%%%

Recently, it was found that interesting effects of frustrated magnetism can also be encountered in the three-dimensional swedenborgite systems. For example, classical $O(3)$ Heisenberg spins on this lattice exhibit a wide spin liquid regime for a certain parameter and temperature range and can undergo a fluctuation driven order by disorder transition to a nematic phase at very low temperatures \cite{Buhrandt2014-2}. It was also shown that the classical Ising model with and without longitudinal magnetic field boasts different phases with extensive and subextensive ground-state degeneracy \cite{Buhrandt2014}. Depending on system parameters, the swedenborgite structure corresponds to either a stack of weakly coupled Kagom\'e layers or an emergent triangular lattice of stiff macro-spins. As a result, one can speculate that the order by disorder of the triangular lattice TFIM and the disorder by disorder of the Kagom\'e TFIM enter into competition with one another. 

In this letter we investigate how the degeneracies of classical Ising model are lifted once quantum fluctuations due to a transverse field are introduced. This is summarized in the phase diagram shown in Fig.~\ref{fig:pd} for the swedenborgite TFIM. For large transverse field values the system will assume the polarized state. The polarized phase is extremely resistant to the in-plane coupling $J_1$ \cite{Powalski2013}. For $J_2 > J_1$ we can map the three-dimensional quantum model to an effective two-dimensional classical model at zero temperature which is formulated in terms of stiff macro-spins living on an emerging triangular lattice. This exotic phase still possesses a subextensively degenerate ground-state manifold corresponding to a macro-spin liquid (MSL). We also find that the phase transition between the polarized phase and the MSL reflects this reduced dimensionality which we deduce from the critical exponents of the transition. 

% Figure 2: Swedenborgite lattice structure
%%%%%%%%%%%%%%%%%%%%%%%%%%%%%%%%%%%%%%%%%%%%%%%%%%%%%%%%%%%%%%%%%%%%%%%%%%%%%%%%%%%%%%%%%%%%%%%%%%%%%%%%%%%%%%%%%%%%%%%%%%%%%%%%
\begin{figure}[t!]
	\centering
	\includegraphics[width=0.35\textwidth]{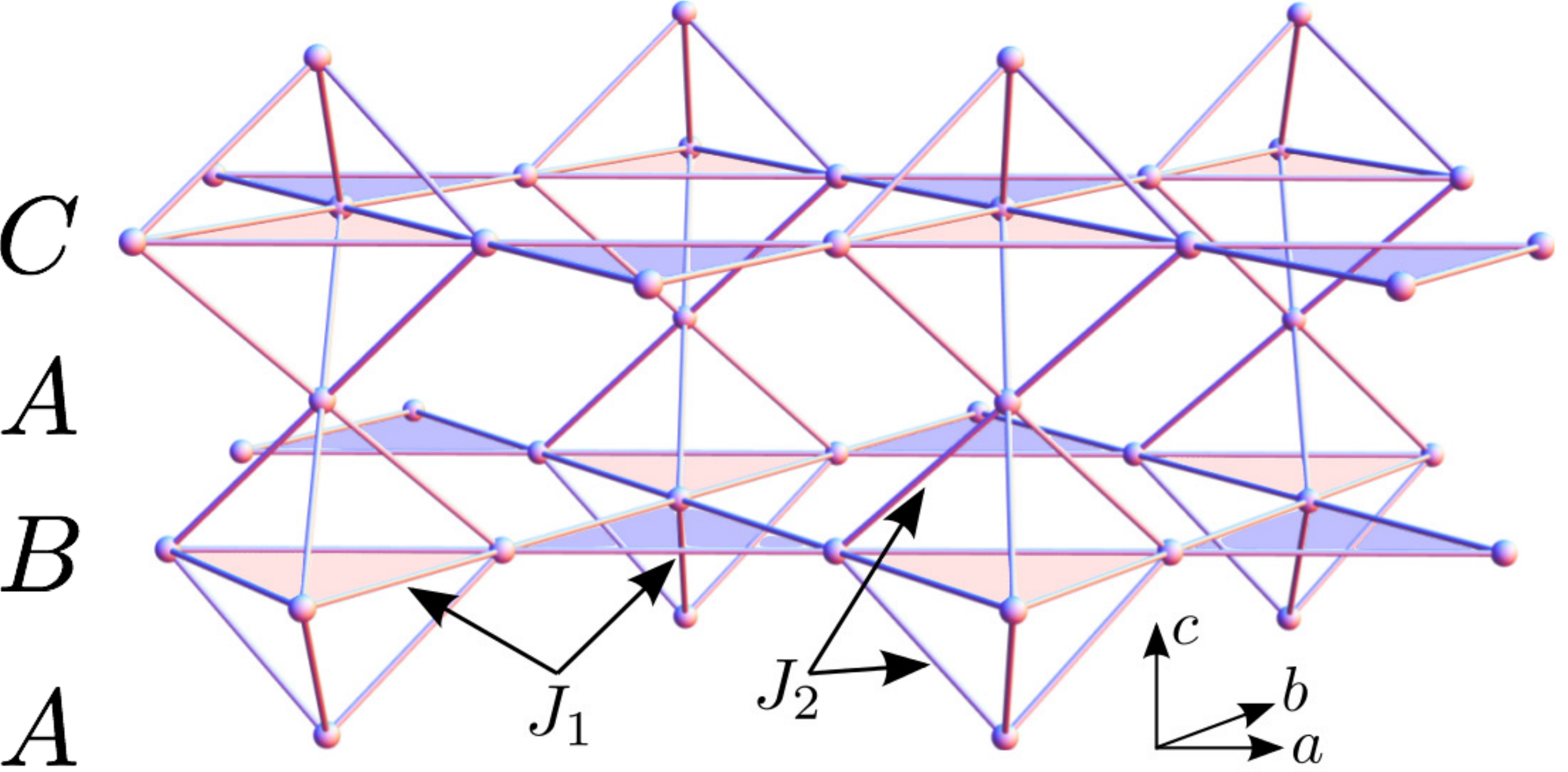}
	\caption{The lattice structure and spin interactions of the swedenborgite lattice. }
	\label{fig:lattice}
\end{figure}
%%%%%%%%%%%%%%%%%%%%%%%%%%%%%%%%%%%%%%%%%%%%%%%%%%%%%%%%%%%%%%%%%%%%%%%%%%%%%%%%%%%%%%%%%%%%%%%%%%%%%%%%%%%%%%%%%%%%%%%%%%%%%%%%
{\it{Swedenborgite TFIM:}}
In the swedenborgite lattice, shown in Fig.~\ref{fig:lattice}, alternating layers of inequivalent Kagom\'e planes (B,C) and triangular spins (A) are stacked in a third spatial dimension in an ABAC... pattern. The resultant structure can alternatively be regarded as a collection of bipyramidal clusters that form columns along the $c$-direction and are  connected by intermediate triangles in the $ab$-plane.

We consider a nearest-neighbor Ising model with two distinct antiferromagnetic interactions: $J_1$ inside the Kagom\'e layers and $J_2$ between the Kagom\'e and triangular layers,
\begin{align} \label{eq: SwedenborgiteHamiltonian}
	\mathcal{H} = J_1\sum_{ {\langle i,j \rangle \in \atop \text{same layer}}} \sigma^x_i \sigma^x_j + J_2\sum_{{\langle i,j \rangle \in \atop \text{diff. layer}}} \sigma^x_i \sigma^x_j + \Gamma \sum_i \sigma^z_i,
\end{align}
with Ising spins $\sigma_i = \pm 1$ and a transverse magnetic field term controlled by parameter $\Gamma$. 

\vspace{12pt}

\begin{table} [h!]
	\centering
	\begin{tabular}{C{1.5cm}|C{6cm}}
		            &	allowed bipyramid configurations \\ \hline 
		$J_2/J_1>1$ &	\includegraphics[width=0.8cm]{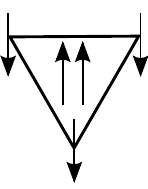} \\
		$J_2/J_1=1$ &	\includegraphics[width=0.8cm]{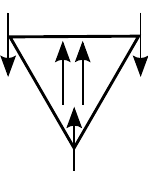}\quad\includegraphics[width=0.8cm]{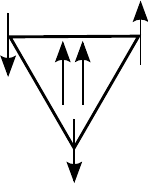}\quad\includegraphics[width=0.8cm]{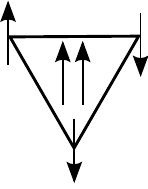}\quad\includegraphics[width=0.8cm]{BP1} \\
		$J_2/J_1<1$ &	\includegraphics[width=0.8cm]{BP2}\quad\includegraphics[width=0.8cm]{BP3}\quad\includegraphics[width=0.8cm]{BP4} \\
	\end{tabular}
	\caption{Optimal bipyramid configurations in the different parameter regimes, where the centre spin pair in each triangle denotes the pre-fixed apical spins of a bipyramid.}
	\label{tab:allBP}
\end{table}

\begin{figure} [t!]
	\includegraphics[width=0.45\textwidth]{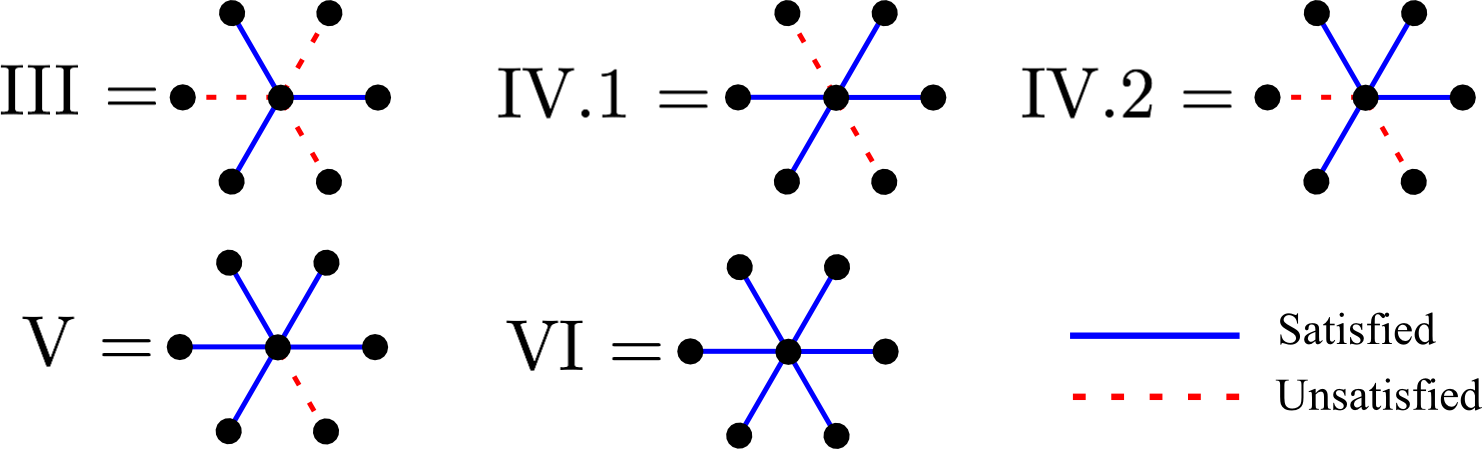}
	\caption{The five different plaquettes allowed on the emergent triangular lattice, up to rotation.}
	\label{fig:plaquettes}
\end{figure}

{\it{Classical limit:}} For $\Gamma=0$, there are two 'spin-liquid' phases ~\cite{Buhrandt2014}, set out in Table~\ref{tab:allBP}, that have distinct degrees of degeneracy. They are separated by a first-order transition, where the transition point has an enlarged degeneracy by virtue of the two manifolds combining in a level crossing. For $J_2>J_1$ the system becomes stiff in $c$-direction, forming macroscopically large spin structures (macro-spins) on the bipyramidal columns. Due to the remaining binary degree of freedom of these macro-spins, the problem reduces in dimension to a version of the classical Ising model on an emergent triangular lattice, formed by the (light blue) intermediate triangles in Fig.~\ref{fig:lattice}. On this lattice, ground states are characterised locally by having two parallel and one antiparallel macro-spin on every subtriangle (2:1 rule). In a bond picture, the ground-state rule dictates that lowest-energy states purely have triangles with precisely two satisfied bonds. As a result the elementary macro-spin units, or plaquettes, allowed on this lattice are restricted to those listed in Fig.~\ref{fig:plaquettes}. 
% Figure 3: Plaquette types of effective dimer model
%%%%%%%%%%%%%%%%%%%%%%%%%%%%%%%%%%%%%%%%%%%%%%%%%%%%%%%%%%%%%%%%%%%%%%%%%%%%%%%%%%%%%%%%%%%%%%%%%%%%%%%%%%%%%%%%%%%%%%%%%%%%%%%%
Furthermore, most combinations of these plaquettes globally disrespect the bond ratio required. Instead, local overcharges of satisfied bonds must be compensated for by including undercharged plaquettes, which means on average we have to have 2:1 satisfied to unsatisfied bonds. There are but three different combinations compatible with this ratio: \mbox{(IV)}, \mbox{(III + V)} and \mbox{(III + III + VI)}.
%%%%%%%%%%%%%%%%%%%%%%%%%%%%%%%%%%%%%%%%%%%%%%%%%%%%%%%%%%%%%%%%%%%%%%%%%%%%%%%%%%%%%%%%%%%%%%%%%%%%%%%%%%%%%%%%%%%%%%%%%%%%%%%%

\begin{figure} [t!]
	\centering
	\includegraphics[width=0.4\textwidth]{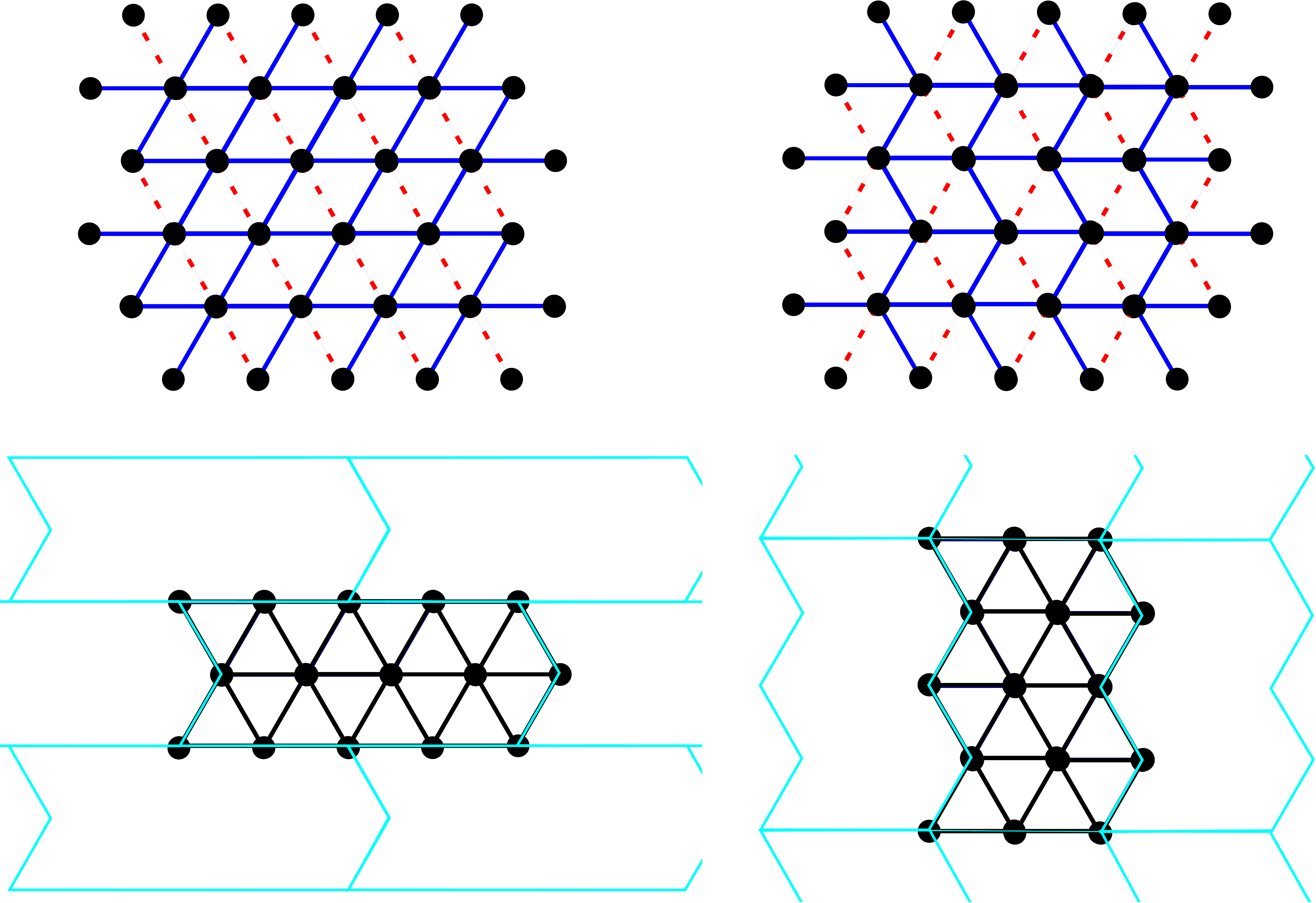}
	\caption{{\it Upper panel}: The minimally (left) and maximally zigzagging (right) stripy structures that form the ground states on the emergent triangular lattice. IV.1-layers will cause straight running stripes, whereas IV.2-layers introduce zigzagging bends. {\it Lower panel}: Clusters used for the exact diaonalization. Grey (cyan) lines denote the unit cells and indicates the chosen periodic boundary conditions.}
	\label{fig: StripyStructures}
\end{figure}

{\it{Quantum-to-classical mapping:}} We consider the effect of the transverse field $\Gamma$ on {\it all} degenerate phases of the classical model individually. For small $\Gamma$, this effect can be studied through a perturbative expansion. For $J_2 > J_1$ off-diagonal processes connecting different ground states are surpressed: in order to connect two degenerate ground states it would be required to flip a whole macro-spin. We consequently only find virtual diagonal processes. This allows to map the three-dimensional quantum model to a two-dimensional classical (no dynamics) dimer model on the triangular lattice with the simple Hamiltonian
\begin{align}
H_{\rm{dim}}=\sum_{i=\text{III,IV,V,VI}} E_i \, | i \rangle \langle i |\quad , 
\label{eq: dimhamil}
\end{align}
where we sum over the different types of plaquettes $i$ with corresponding energy $E_i$ (note that this model still is subject to the 2:1 rule!). We have explicitly determined the potential terms up to fourth order in perturbation theory. To second order in $\Gamma$ we find that the three different plaquette combinations reduce their energy by the same amount. This behaviour changes only at fourth order in $\Gamma$ \cite{Supp}. A relative shift in the energy reduction of the plaquette states results in the hierarchy $E_\text{IV} < \frac{1}{2} \left( E_\text{III} + E_\text{V} \right) < \frac{1}{3} \left(2  E_\text{III} + E_\text{VI} \right)$. It is thus energetically advantageous to populate the lattice with IV-plaquettes. This type of plaquette still comes in two distinct varieties. The geometry of these forms causes the system to crystallise in one direction, leaving only a single dimension undetermined. In this direction `stripy' configurations with a varying degree of zigzagging will make up the ground states. The two extremal cases of such configurations are depicted in Fig.~\ref{fig: StripyStructures}, but any combination of the two is allowed resulting in a subextensively degenerate ground-state manifold. These stripy lattice coverings have also been suggested as ground-state configuration for a model with antiferromagnetic interactions between Ising spins on an elastic triangular lattice ~\cite{Shokef2011}. 

It is natural to ask whether subextensive degeneracy is lifted beyond order four shown in Eq.~\eqref{eq: FourthOrderEnergyContributions}. From a perturbative perspective the degeneracy can be traced back to the topological equivalence of graphs IV.1 and IV.2. The degeneracy is lifted when virtual processes of the minimally stripy structure have no symmetric analogue for the maximally zigzagging stripy structure. Similar to Ref.~\cite{Powalski2013}, this can only occur in our situation when fluctuations on loops of the Kagom\'e lattice play a role. The relevant minimal loop on the Kagom\'e lattice is a hexagon consisting of six spins. In each order the perturbation leads to local spin flips linking two neighboring bonds of a hexagon. Consequently, up to order ten there is no linked process which covers the hexagon and the subextensive degeneracy is not lifted. In fact, the system seems to be even more reluctant to order in either fashion which we have checked by exact diagonalization of the two periodic clusters shown in the lower panel of Fig.~\ref{fig: StripyStructures}. Each cluster consists of 24 spins in the Kagom\'e plane of the original swedenborgite lattice and is constructed such that both, the minimally and maximally zigzagging, stripy structure fits on the cluster with the same loop length. Note that the spins above and below the displayed Kagom\'e plane are chosen in a frozen configuration for the exact diagonalization. Surprisingly, the ground-state energy of both stripy structures is exactly degenerate on both clusters up to numerical precision implying that all virtual fluctuations up to infinite order, which fit on the shown clusters, do not lift the extensive degeneracy and the system remains disordered. We consequently conclude that this systems provides an example of a macro-spin liquid.     
% Figure 4: Minimally and maximally zigzagging stripy structures
%%%%%%%%%%%%%%%%%%%%%%%%%%%%%%%%%%%%%%%%%%%%%%%%%%%%%%%%%%%%%%%%%%%%%%%%%%%%%%%%%%%%%%%%%%%%%%%%%%%%%%%%%%%%%%%%%%%%%%%%%%%%%%%%
%%%%%%%%%%%%%%%%%%%%%%%%%%%%%%%%%%%%%%%%%%%%%%%%%%%%%%%%%%%%%%%%%%%%%%%%%%%%%%%%%%%%%%%%%%%%%%%%%%%%%%%%%%%%%%%%%%%%%%%%%%%%%%%%

Notice that this analysis breaks down for $J_2 \leq J_1$, where some of the different spin bipyramid configurations in the degenerate ground-state manifold are connected through a second order spin flip process. There are off-diagonal tunneling terms between the distinct classical ground states that impair any attempt at classical inspection of the problem. 

{\it{High-field expansion:}} For $\Gamma \gg J_1,J_2$, we can understand the excitations generated by a single spin flip as quasiparticles above a vacuum provided by the fully polarized ground state. We introduce hardcore boson operators $(a^\dagger, a)$ acting on Kagom\'e plane sites and $(b^\dagger, b)$ operating in the triangular layers which generate or destroy a spin flip. The Hamiltonian~\eqref{eq: SwedenborgiteHamiltonian} expressed in terms of these operators for $\Gamma=1/2$ reads
\begin{eqnarray}
 \label{eq: QuasiparticleHamiltonian}
 \Hamil &=& -\frac{N}{2}+\sum_{i,\mu\in a,b} \hat{n}_i^{(\mu)}+ J_1 \sum_{\langle i,j \rangle} \left( a_i^\dagger a_j^\dagger + a_i^\dagger a^{\phantom{\dagger}}_j + \text{h.c.} \right)\nonumber\\
                         && + J_2 \sum_{\langle i,j \rangle} \left( a_i^\dagger b_j^\dagger + a_i^\dagger b^{\phantom{\dagger}}_j + \text{h.c.} \right)\nonumber\\
                         &=&\Hamil_0 + \sum_{\nu = 1,2} J_\nu \left( T_0^{(\nu)} + T_{+2}^{(\nu)} + T_{-2}^{(\nu)} \right)
\end{eqnarray}
where $\Hamil_0 = -N/2+ \sum_i \left( a_i^\dagger a^{\phantom{\dagger}}_i + b_i^\dagger b^{\phantom{\dagger}}_i \right)$, $N$ is the total number of sites, and the index $m = \{0, \pm 2\}$ refers to the change of the quasiparticle number due to the operator $T_m^{(\nu)}$.

A pCUT (perturbative continuous unitary transformation) \cite{Knetter2000,Knetter2003} transforms Hamiltonians of the form Eq.~\eqref{eq: QuasiparticleHamiltonian} into effective quasiparticle conserving Hamiltonians $\mathcal{H}_{\rm eff}$ so that $[\mathcal{H}_{\rm eff},\mathcal{H}_0]=0$. This first step of the method can be done model-independently, i.e., the effective Hamiltonian in the thermodynamic limit is calculated perturbatively up to high orders. The second model-specific step is to normal order $\mathcal{H}_{\rm eff}$ which is done most efficiently via a linked-cluster expansion. The computational effort of a linked-cluster expansion scales exponentially with the number of graphs as well as with the number of perturbative parameters, implying that the three-dimensional model under consideration with two parameters $J_\nu$ ($\nu=1,2$) is of challenging complexity. Fortunately, the recently introduced white-graph expansion \cite{Coester2015} is perfectly suited for this problem. Here we concentrate on the physical properties of a single quasiparticle in order to investigate the breakdown of the polarized high-field phase. Applying a white-graph expansion, we determine the effective one quasiparticle Hamiltonian up to order 11 in both parameters $J_\nu$.  

\begin{figure} [t!]
	\includegraphics[width=0.4\textwidth]{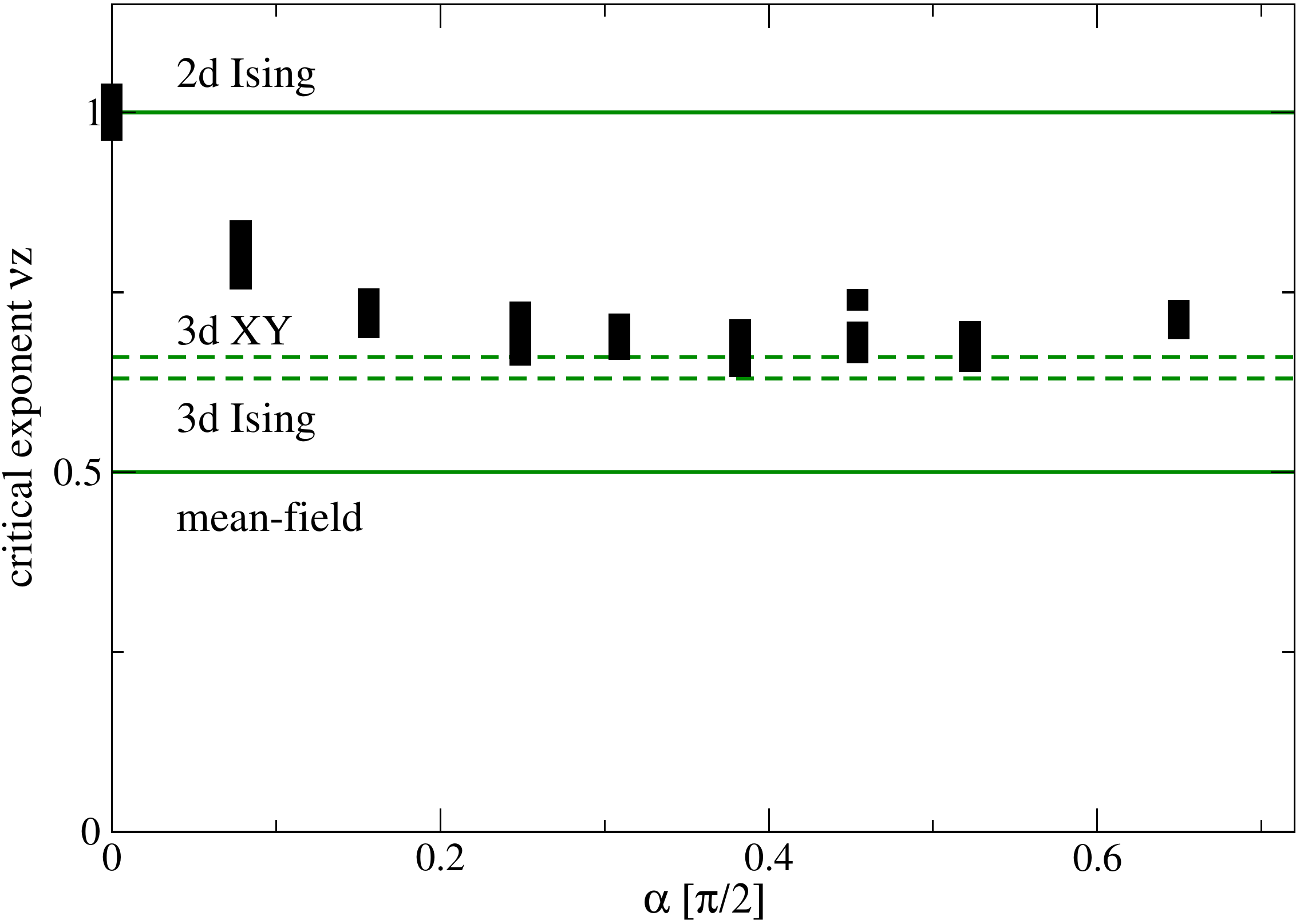}
	\caption{Critical exponent $z\nu$ as a function of $\alpha$ where $\tan \alpha=J_1 / J_2$ obtained from various Dlog-Pad\'{e} extrapolations of the high-field gap $\Delta (J_1,J_2)$. Solid and dashed horizontal lines indicate known critical exponents for certain universality classes.}
	\label{fig:crit_exponent}
\end{figure}

The effective one-particle hopping Hamiltonian can then be block-diagonalized via Fourier transformation with respect to the eight-site unit cell of the lattice. One obtains eight one-particle bands $\omega_{n}(\vec{k})$ with $n\in\{1,\ldots,8\}$ and $\vec{k}=(k_a,k_b,k_c)$. Here $k_a$ and $k_b$ are the momenta in the Kagom\'e $ab$-planes while $k_c$ is parallel to the $c$-direction. We find that the one-particle gap $\Delta$ is always located at $\vec{k}=(2\pi/3,-2\pi/3,0)$ (and other momenta related by symmetry) fully consistent with the findings for the pure Kagom\'e TFIM \cite{Powalski2013}. This momentum is not compatible with any configuration which can be constructed from the plaquettes discussed in Fig.~\ref{fig: StripyStructures}. Next we use Dlog-Pad\'{e} extrapolations \cite{Guttmann89} for the one-particle gap $\Delta (J_1,J_2)$ to estimate the location of quantum critical points $(J_1^{\rm c},J_2^{\rm c})$ with $\Delta (J_1^{\rm c},J_2^{\rm c})=0$ and the corresponding critical exponent $z\nu$. The results for the phase diagram are displayed in Fig.~\ref{fig:pd} and for the critical exponent in Fig.~\ref{fig:crit_exponent}.  
% Figure 5: Critical exponent z\nu 
%%%%%%%%%%%%%%%%%%%%%%%%%%%%%%%%%%%%%%%%%%%%%%%%%%%%%%%%%%%%%%%%%%%%%%%%%%%%%%%%%%%%%%%%%%%%%%%%%%%%%%%%%%%%%%%%%%%%%%%%%%%%%%%%
%%%%%%%%%%%%%%%%%%%%%%%%%%%%%%%%%%%%%%%%%%%%%%%%%%%%%%%%%%%%%%%%%%%%%%%%%%%%%%%%%%%%%%%%%%%%%%%%%%%%%%%%%%%%%%%%%%%%%%%%%%%%%%%%

In the pure Kagom\'e limit, $J_2=0$, we recover the results of Ref.~\onlinecite{Powalski2013}. The lowest band is completely flat up to and including order seven, and a specific momentum is only chosen from order eight. Extrapolating the one-particle gap for small values of $J_2$ gives no indications for a gap closing at any values of $(J_1,J_2)$. Let us mention that the extrapolations might not be very accurate for $|(J_1,J_2)|\rightarrow\infty$ and no complementary calculation for the limit $\Gamma\rightarrow 0$ can be done efficiently for the 3d swedenborgite lattice in contrast to the two-dimensional counterpart on the Kagom\'e lattice \cite{Powalski2013}. Nevertheless, our findings are consistent with a remarkable disorder by disorder scenario in three dimensions for an extended parameter range including the pure Kagom\'e TFIM.

The physics is fundamentally different in the regime $J_2>J_1$. In the limit $J_1=0$ one has isolated and unfrustrated TFIMs on the bipyramidal chains along the $c$-direction. One therefore expects a quantum phase transition in the 2D Ising universality class with $z\nu=1$ for these one-dimensional quantum systems which is quantitatively confirmed by our calculation. We find \mbox{$(0,J_2^{\rm c})=(0,0.239\pm 0.002)$} and \mbox{$z\nu=1.00\pm 0.02$}. Introducing a finite $J_1$ we deduce a non-trivial quantum critical line. For small values of $J_1$, the critical value of $J_2$ is decreased, i.e., quantum fluctuations due to the frustrated Ising interactions in the Kagom\'e planes weaken the polarized phase. Interestingly, this behaviour changes for larger ratios $J_1/J_2$ meaning the polarized phase is stabilized by quantum fluctuations in this part of the phase diagram. This quantum critical line shows a remarkable behaviour of the corresponding critical exponents: the critical exponent $z\nu$ is inconsistent with an expected value in (3+1) dimensions, but one observes for a broad range of ratios $J_1/J_2$ a value $\approx 0.7$ typical for (2+1) dimensional criticality. This suggests that the dimensional reduction observed in the limit $J_2>J_1$ also is present in the quantum critical properties.         

{\it{Discussion}} One may wonder how our findings for the MSL and the polarized phase concerning the nature of the quantum phase transition can be reconciled. In the simplest scenario there is a first-order phase transition between these two phases, since the momentum of the one-particle gap in the high-field phase is not compatible with any ground state configuration in the MSL phase. Let us mention that a single series expansion in one phase is not able to detect first-order phase transitions. However, the lack of attractive quasiparticle interactions in the polarized phase gives no obvious tendency towards such a direct transition. In the other scenario there is a quantum critical line between the two phases as suggested by the high-order series expansion. This is clearly more interesting. In this case the physics must be even more exotic, since an intermediate phase is expected to emerge as a breakdown of the MSL. Keeping in mind the extremely small energy scales involved, an accurate description of this intermediate phase as well as the attached quantum phase transitions is certainly a formidable challenge.     

To conclude, our study reveals the fascinating physics of highly frustrated quantum magnets in three spatial dimensions. The swedenborgite lattice as a prototypical system displays several exotic features due to the competition between geometric frustration and quantum fluctuations giving rise to a rich phase diagram. In a broader perspective, the emergence of macro-spins as classical entities of three-dimensional frustrated quantum systems deserves more investigations in future research since it appears to give a route towards exotic quantum criticality where the critical dynamics is strongly constrained and direction dependent.   

{\it{Acknowledgement:}}
We acknowledge funding from the DFG FR 2627/3-1 (S.B. and L.F.). This work is part of the D-ITP consortium, a program of the Netherlands Organisation for Scientific Research (NWO) that is funded by the Dutch Ministry of Education, Culture and Science (OCW) (T.S.S. and L.F.).

\newpage

\newpage

\clearpage

\begin{widetext}
\begin{center}
\textbf{\large Supplemental Material}
\end{center}
\end{widetext}
\setcounter{equation}{0}
\setcounter{figure}{0}
\setcounter{table}{0}
\setcounter{page}{1}
\makeatletter
\renewcommand{\theequation}{S\arabic{equation}}
\renewcommand{\thefigure}{S\arabic{figure}}
\renewcommand{\bibnumfmt}[1]{[S#1]}
\renewcommand{\citenumfont}[1]{S#1}
\section{Effective dimer model}
The fourth-order energy shifts of the effective dimer model read 
\begin{align} \label{eq: FourthOrderEnergyContributions}
\Delta E_\text{III}^{(4)}	&= \frac{3 (2J_2 + J_1)}{64 (J_2-J_1)^3 (2J_2-J_1)}, \nonumber \\
\Delta E_\text{IV}^{(4)}	&= \frac{{J_1}^2 + 6 {J_2}^2 - 3 J_1 J_2}{64 (J_2-J_1)^3 J_2 (2J_2-J_1)}, \\
\Delta E_\text{V}^{(4)}		&=  - \frac{ 3 {J_1}^4 - 6 {J_2}^4 - 7 {J_1}^3 J_2 + 4 {J_1}^2 {J_2}^2 + 3 J_1 {J_2}^3}{64 (J_2 - J_1)^3 {J_2}^3 (2J_2 + J_1)}, \nonumber \\	
\Delta E_\text{VI}^{(4)}	&= \frac{3 (2J_2 + 3 J_1)}{64 {J_2}^3 (2J_2+J_1)}. \nonumber
\end{align}

\section{Extrapolation of the breakdown of one-particle gap}

\begin{figure}[h]
\includegraphics[width=0.4\textwidth]{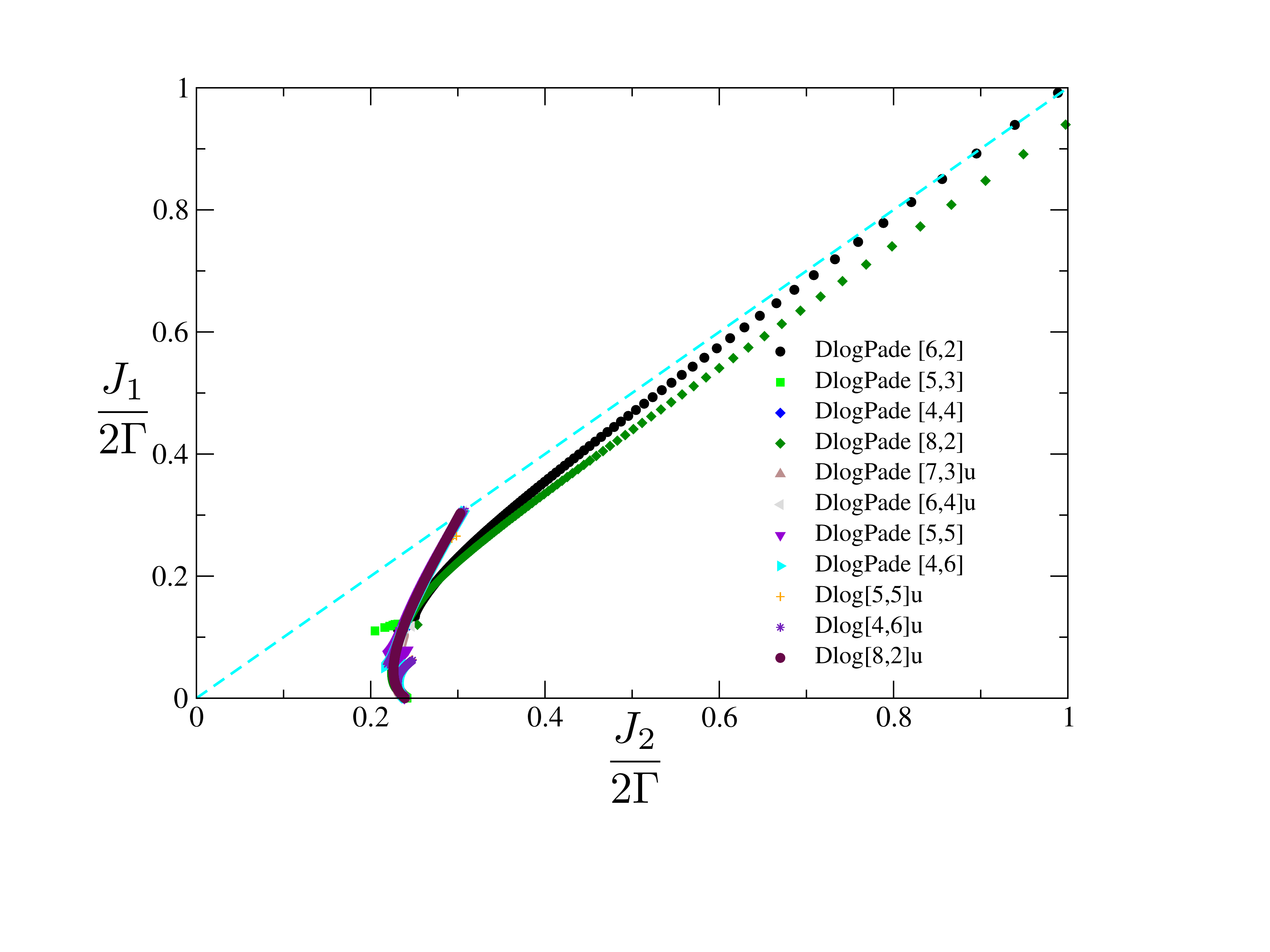}
\caption{Phase boundary using different Dlog-Pad\'{e} extrapolation schemes. For small values of $2J_1/\Gamma$ we find very consistent picture. }\label{fig:dlogpade}
\end{figure}

We have used different Dlog-Pad\'{e} extrapolations for the one-particle gap $\Delta (J_1,J_2)$ to estimate the location of quantum critical points $(J_1^{\rm c},J_2^{\rm c})$ with $\Delta (J_1^{\rm c},J_2^{\rm c})=0$. The results for the phase boundary are shown in Fig.~\ref{fig:dlogpade}

\end{document}